\documentclass[a4paper,UKenglish,cleveref, autoref, thm-restate]{lipics-v2021}
\usepackage{booktabs}
\usepackage{relsize}

\title{Enriching Location Representation with Detailed Semantic Information} 

\author{Junyuan Liu}{SpaceTimeLab, University College London, UK \and \url{}}{junyuan.liu.22@ucl.ac.uk}{https://orcid.org/0009-0009-7194-6868}{}

\author{Xinglei Wang}{SpaceTimeLab, University College London, UK}{xinglei.wang.21@ucl.ac.uk}{https://orcid.org/0000-0002-9824-7663}{}

\author{Tao Cheng\footnote{Corresponding author}}{SpaceTimeLab, University College London, UK}{tao.cheng@ucl.ac.uk}{https://orcid.org/0000-0002-5503-9813}{}

\authorrunning{J. Liu, X. Wang, and T. Cheng} 

\Copyright{Junyuan Liu, Xinglei Wang, and Tao Cheng} 



\nolinenumbers 

\EventEditors{Katarzyna Sila-Nowicka, Antoni Moore, David O'Sullivan, Benjamin Adams, and Mark Gahegan}
\EventNoEds{5}
\EventLongTitle{13th International Conference on Geographic Information Science (GIScience 2025)}
\EventShortTitle{GIScience 2025}
\EventAcronym{GIScience}
\EventYear{2025}
\EventDate{August 26--29, 2025}
\EventLocation{Christchurch, New Zealand}
\EventLogo{}
\SeriesVolume{346}
\ArticleNo{3}

\begin{document}

\maketitle

\begin{abstract}
Spatial representations that capture both structural and semantic characteristics of urban environments are essential for urban modeling. Traditional spatial embeddings often prioritize spatial proximity while underutilizing fine-grained contextual information from places. To address this limitation, we introduce \textbf{CaLLiPer+}, an extension of the CaLLiPer model that systematically integrates Point-of-Interest (POI) names alongside categorical labels within a multimodal contrastive learning framework. We evaluate its effectiveness on two downstream tasks—land use classification and socioeconomic status distribution mapping—demonstrating consistent performance gains of 4\% to 11\% over baseline methods. Additionally, we show that incorporating POI names enhances location retrieval, enabling models to capture complex urban concepts with greater precision. Ablation studies further reveal the complementary role of POI names and the advantages of leveraging pretrained text encoders for spatial representations. Overall, our findings highlight the potential of integrating fine-grained semantic attributes and multimodal learning techniques to advance the development of urban foundation models.
\end{abstract}
\section{Introduction}
\label{sec:introduction}

Spatial representations form the backbone of urban analysis, serving as essential tools for understanding and modeling complex urban systems. They underpin various applications, including urban functional distribution mapping \cite{huang2022sppe,huang2023hgi}, land use classification \cite{jean2019tile2vec}, socioeconomic indicator estimation \cite{jean2016combining}, future visitor prediction \cite{feng2017poi2vec}, and next-location prediction \cite{hong2023context}. Traditional approaches typically encode locations as numeric coordinates or rely on spatial proximity \cite{mai2020iclr,mai2023sphere2vec,yao2017sensing}, effectively capturing physical distance and structure. However, they often fail to capture the intricate functional interdependencies between places that drive urban dynamics.

In contrast, “platial” concepts emphasize the additional layers of meaning that humans ascribe to spaces, interpreting them through social, cultural, and functional attributes \cite{goodchild2020platial}. Point-of-Interest (POI) data offers a practical entry point for these attributes, as it couples spatial coordinates with descriptive names and labels. Such semantic information elucidates how different places function and interact within the broader urban landscape. Nevertheless, many existing embedding methods continue to emphasize spatial distance or simple categorical labels \cite{huang2022sppe,huang2023hgi,yan2017itdl,yao2017sensing,zhai2019beyond}, underutilizing POI data’s finer-grained insights.

Recent innovations in deep learning and natural language processing \cite{reimers2019sbert,openai2022chatgpt,dubey2024llama3} facilitate richer semantic alignments within spatial data. Notably, multimodal contrastive learning \cite{radford2021clip} has proven effective in aligning geographic coordinates with textual descriptions, thereby enhancing the semantic depth of spatial embeddings. A prime example is CaLLiPer \cite{wang2025multi}, which aligns POI types with spatial coordinates to yield improvements in downstream tasks. However, CaLLiPer treats POI types as broad categorical labels, potentially overlooking the granular detail contained in POI names. Such names often provide specific and context-rich information, ranging from “Starbucks Coffee” to “John’s Hardware Store,” which can further enrich location understanding and distinguish unique POIs. Yet, the systematic integration of POI names into general-purpose spatial embeddings through multimodal contrastive learning remains underexplored. Addressing this gap is crucial for fully capturing the nuanced semantics of urban environments and advancing more comprehensive urban representation models.

To enhance spatial embeddings with richer semantic detail, we incorporate POI names alongside type labels into a multimodal contrastive learning framework. Building on the original CaLLiPer model, we propose an extended version called CaLLiPer+. We evaluate its effectiveness in two downstream tasks—Land Use Classification (LUC) and Socioeconomic Status Distribution Mapping (SDM)—as well as in an additional location retrieval task.

Our contributions are as follows:
\begin{enumerate}
    \item We extend the CaLLiPer framework by incorporating POI names alongside type information, resulting in a unified model, CaLLiPer+ (§\ref{sec:methodology}).
    \item We evaluate the enriched semantic representation on two downstream tasks, showing consistent performance gains of 4\% to 11\% over POI-type-only models (§\ref{subsec:results_downstream}).
    \item We conduct retrieval experiments to assess the model’s ability to capture urban concepts, and show that enriched semantics and advanced text encoders lead to better conceptual understanding (§\ref{subsec:retrieval}).
    \item We demonstrate the effectiveness of contrastive learning with a pretrained encoder for location representation, and highlight the potential of the resulting embeddings for downstream applications (§\ref{sec:discussion}).
\end{enumerate}

\section{Related Work}
\label{sec:relatedwork}

\subsection{Word Embeddings and Sentence Embeddings}

The advancement of natural language processing (NLP) has led to powerful embedding techniques that transform textual data into high-dimensional vector spaces, enabling machines to better process and understand linguistic semantics. Early word embedding models such as Word2Vec \cite{mikolov2013efficient} and GloVe \cite{pennington2014glove} revolutionized NLP by capturing semantic relationships between words based on their co-occurrence in large text corpora.

Building on these foundational methods, sentence embedding models like Sentence-BERT \cite{reimers2019sbert} and SimCSE \cite{gao2021simcse} were developed to generate dense representations of entire phrases or sentences while preserving contextual nuances. More recently, large language models (LLMs) such as BERT \cite{devlin2019bert}, GPT \cite{radford2019language, brown2020language, openai2023gpt4}, and LLaMA \cite{touvron2023llama} have further enhanced text embedding capabilities, facilitating sophisticated semantic extraction across various textual contexts, including POI descriptions and names.

These advancements in NLP offer new opportunities to incorporate linguistic semantics into geospatial models, enabling the embedding of POI names and descriptions to enrich spatial representations beyond purely numerical features.

\subsection{Spatial Embeddings with POIs}

Spatial embedding techniques aim to encode geographic entities into vector spaces, capturing their spatial and functional relationships. POI data, which contains both geographic coordinates and semantic attributes, has been widely utilized in urban studies for tasks such as land use classification, urban function recognition, and socioeconomic mapping.

Early approaches to spatial embeddings primarily leveraged POI categories to model urban entity co-occurrence. Yao et al. \cite{yao2017sensing} proposed a method that traversed POIs within a geographic region using shortest-path algorithms to extract co-occurrence patterns. Place2Vec \cite{yan2017itdl} applied a K-nearest neighbor (KNN) sampling strategy with distance decay to model spatial proximity, while Doc2Vec \cite{niu2021delineating} treated urban regions as documents and POIs as words, learning region embeddings based on the co-occurrence of POI categories within predefined spatial boundaries. These methods effectively captured the functional composition of urban spaces but treated POIs as categorical variables, overlooking their individual characteristics and richer semantic meanings.

To provide more distinguishing information for individual POIs, recent methods have explored integrating additional semantic attributes into spatial embeddings. Huang et al. \cite{huang2022sppe} introduced the Semantics-Preserved POI Embedding (SPPE) model, which incorporates both spatial co-occurrence patterns and categorical semantics to enhance the representation of POI distributions. Similarly, HGI \cite{huang2023hgi} employed hierarchical graph-based embeddings to capture multi-level semantic relationships among POIs, urban regions, and cities. While these methods improved the semantic richness of spatial representations, they still primarily rely on categorical classifications and predefined spatial structures, limiting their adaptability to diverse urban environments.

Existing methods for spatial embeddings primarily aggregate POI information within predefined regions or construct complex spatial contexts to infer urban functions. These approaches often rely on indirect or coarse-grained representations. With the growing availability of detailed POI datasets and advances in NLP, a more direct and efficient approach is to embed individual POIs by leveraging their inherent semantic information, such as names, which provide fine-grained functional and cultural context.

\subsection{Multimodal Contrastive Learning for Geospatial Data}

Multimodal contrastive learning has recently gained traction as an effective method for aligning heterogeneous data sources, enabling the integration of spatial coordinates with diverse information. This approach leverages contrastive objectives to maximize similarity between aligned data pairs (e.g., a location and its textual description) while distinguishing them from unrelated samples.

UrbanCLIP \cite{yan2024urbanclip} proposed a pre-training approach for urban region representation by generating textual descriptions for satellite images using large language models and training an image encoder via a CLIP-like framework. Similarly, GeoCLIP \cite{vivanco2024geoclip} and SatCLIP \cite{klemmer2023satclip} extended contrastive learning to geospatial data by aligning satellite imagery with geographic coordinates, supporting tasks such as geo-localization and environmental monitoring. The CaLLiPer model \cite{wang2025multi} advanced this concept by aligning POI type semantics with spatial coordinates through multimodal contrastive learning, demonstrating improved performance in land use classification and socioeconomic status mapping.

Despite these advances, existing models primarily focus on solely POI type or complex visual data, overlooking the potential benefits of simply incorporating distinguishing semantics of POI names into contrastive learning settings, which contain rich, context-specific information that can enhance the semantic depth of spatial embeddings, offering more nuanced insights into urban functions and structures. The underutilization of POI names in multimodal frameworks is still a significant gap in current geospatial representation learning research.

\section{Methodology}
\label{sec:methodology}

\subsection{Overview}

This study builds upon the CaLLiPer framework \cite{wang2025multi}, a multimodal contrastive learning model designed to align spatial coordinates with semantic information extracted from POI data. While the core architecture remains consistent with CaLLiPer, we introduce a key modification: the integration of POI names into the textual descriptions, enriching the semantic representation of urban spaces.

\begin{figure}[t]
    \centering
    \includegraphics[width=1\linewidth]{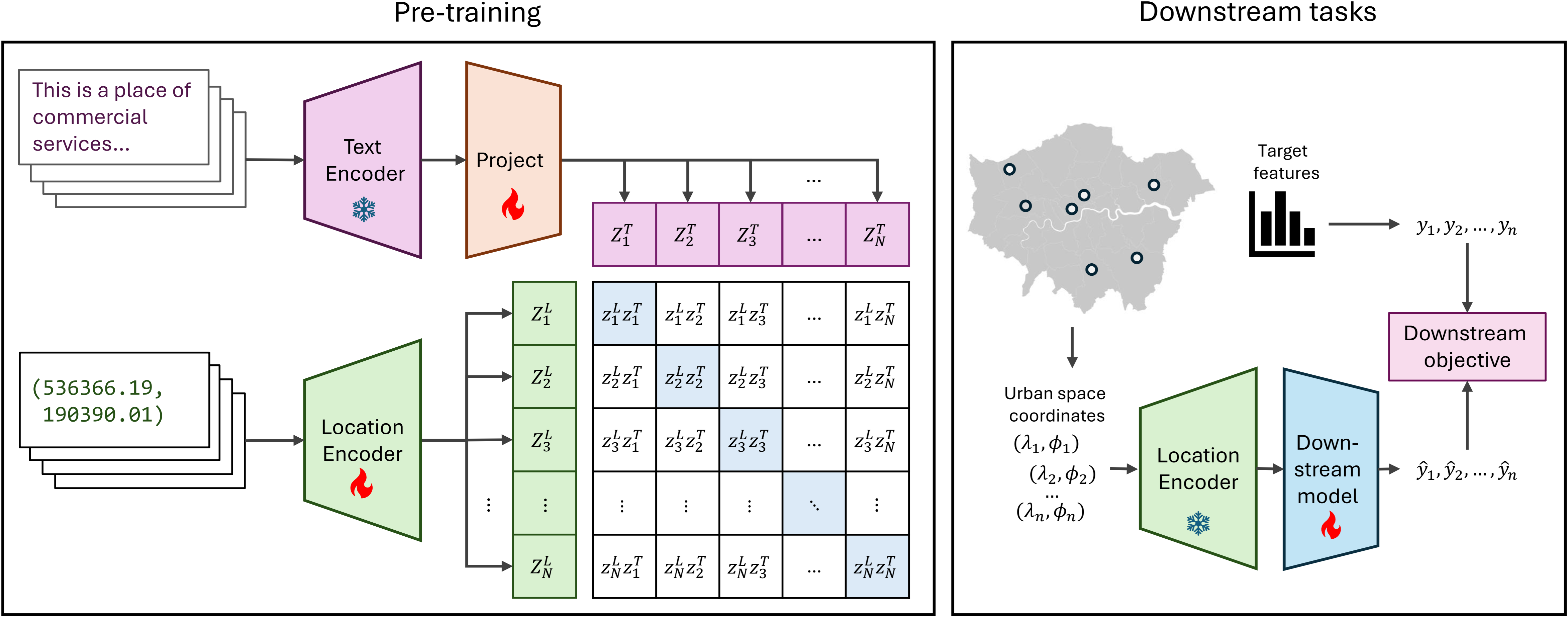}
    \caption{Architecture of the CaLLiPer+ model \cite{wang2025multi}. POI names are incorporated into the textual descriptions processed by the text encoder, enhancing the semantic richness of the spatial embeddings.}
    \label{fig:model_architecture}
\end{figure}

Figure~\ref{fig:model_architecture} illustrates the overall architecture, which consists of three key components: a location encoder, a text encoder, and a projection layer. These components are jointly optimized using a contrastive learning objective to align spatial and semantic information effectively.

\textbf{Location encoder.} The location encoder maps spatial coordinates into a continuous vector space. It applies a positional encoding function to transform raw geographic coordinates into structured representations, followed by a fully connected neural network to generate location embeddings. In this work, we apply the Grid \cite{mai2020iclr} positional encoding function.

\textbf{Text encoder.} The text encoder is a frozen pretrained embedding model, such as Sentence-BERT \cite{reimers2019sbert}, LLaMA \cite{touvron2023llama}, or GPT \cite{openai2023gpt4}, which generates semantic embeddings from the enriched POI descriptions. By incorporating POI names alongside categorical information, it captures more nuanced semantic details, improving the discriminative power of the embeddings.

\textbf{Projection layer.} To facilitate direct comparison between spatial and textual embeddings, a linear projection layer maps both of them into a common vector space of dimension $d$. This projection ensures compatibility between modalities, enabling effective contrastive learning.

\subsection{Enriching POI Descriptions with Names}

In the original CaLLiPer model, POI semantics are represented solely by two levels of categorical labels from the Ordnance Survey. While effective for generalizing urban functions, this approach overlooks the rich, context-specific information embedded in POI names. Names often convey distinctive characteristics, such as cultural significance, brand identity, or specialized services, which are not captured by generic type labels. For instance, ``McDonald's'' may evoke a different functional connotation compared to a generic ``restaurant,'' particularly in terms of cuisine style or consumption level.

To address this limitation, we extend the POI descriptions by integrating names directly into the semantic representation. For each POI $p_i$, we construct a combined description $d_i$ that incorporates the name $n_i$, the first-level category $t_{1i}$, and the second-level class $t_{2i}$ using a templated format designed to enhance the model's understanding of the spatial context:

\begin{equation}
    d_i = \text{Template}(n_i, t_{1i}, t_{2i}) = \text{``A place of } [t_{2i}] \text{, a type of } [t_{1i}] \text{, named } [n_i] \text{.''}
\end{equation}

This enriched template ensures that the text encoder can capture both general category information and the specific nuances associated with individual POIs. By incorporating POI names, the model captures finer-grained semantic details that improve its ability to differentiate between places within the same category. This includes recognizing brand prestige (e.g., “Hilton Hotel” vs. “Budget Inn”), specific function within the same type (e.g., “The British Museum” vs. “National Gallery”), and scale or exclusivity (e.g., “local farm market” vs. “Harrods”). This richer semantic embedding enhances the model’s capacity to represent the diversity and complexity of urban environments more accurately.

\subsection{Multimodal Contrastive Learning Framework}

The multimodal contrastive learning framework aligns spatial coordinates with detailed textual semantics in a shared embedding space. The goal is to ensure that a POI’s spatial representation is closely aligned with its semantic description, while unrelated pairs are pushed apart.

Each POI is represented by two embeddings:
\begin{align}
    z_i^{(s)} &= f_s(\mathbf{x}_i) \quad \text{(spatial embedding)} \\
    z_i^{(p)} &= W_t f_t(d_i) \quad \text{(textual embedding with name and type)}
\end{align}

where $f_s$ is the spatial encoder that transforms the geographic coordinates $\mathbf{x}_i$ into a vector representation, and $f_t$ is a pretrained text encoder that processes the enriched POI descriptions $d_i$, followed by a projection layer $W_t$ to align the dimension with spatial embedding. The inclusion of POI names in $d_i$ ensures that the text embeddings capture both high-level categorical information and fine-grained, context-specific details.

\textbf{Contrastive learning objective.} The alignment between spatial and textual embeddings is achieved using the InfoNCE loss \cite{radford2021clip}, which encourages positive pairs (i.e., a POI's location and its enriched description) to be similar, while pushing apart negative pairs (i.e., mismatched locations and descriptions). The loss is defined as:

\begin{equation}
    \mathcal{L} = -\frac{1}{2N} \left[\sum_{i=1}^{N} \log \frac{\exp(z_i^{(s)} \cdot z_i^{(p)} / \tau)}{\sum_{j=1}^{N} \exp(z_i^{(s)} \cdot z_j^{(p)} / \tau)} 
    + \sum_{i=1}^{N} \log \frac{\exp(z_i^{(p)} \cdot z_i^{(s)} / \tau)}{\sum_{j=1}^{N} \exp(z_i^{(p)} \cdot z_j^{(s)} / \tau)} \right],
\end{equation}

where $\cdot$ denotes cosine similarity between embeddings, and $\tau$ is a temperature parameter that controls the sharpness of the distribution. This symmetric loss is applied to both spatial-to-textual and textual-to-spatial alignment, ensuring consistent alignment of embeddings from both modalities.

\textbf{Advantages of enriched semantics.} Incorporating POI names into the contrastive framework enhances the model's ability to capture fine-grained urban semantics. The enriched descriptions provide the following benefits:
\begin{itemize}
    \item Improved discrimination: The model can better differentiate between places within the same category by leveraging unique names.
    \item Context awareness: Names often imply cultural, historical, or functional context, enriching the model's understanding of urban environments.
    \item Enhanced transferability: The enriched embeddings generalize better across diverse tasks.
\end{itemize}

In summary, our approach enhances the original CaLLiPer framework by incorporating POI names into the textual descriptions, leading to richer, more discriminative spatial embeddings through multimodal contrastive learning.

\section{Experiments}
\label{sec:experiments}

\subsection{Experimental Setup}

To evaluate the impact of incorporating POI names into the spatial-semantic embeddings, we conducted experiments on two urban analytics tasks: Land Use Classification (LUC) and Socioeconomic Status Distribution Mapping (SDM). Additionally, we performed location retrieval to observe the model’s ability to capture high-level urban concepts.

\subsection{Datasets}

\textbf{Point-of-Interest data.} We use POI data from the Ordnance Survey via Digimap \footnote{\url{https://digimap.edina.ac.uk/}}, covering the Greater London area. The dataset contains approximately 340,000 POIs, each with geographic coordinates, a name, and categorical labels. POIs are classified into a hierarchical taxonomy. These data provide detailed spatial and semantic insights into London's urban environment.

\textbf{Land use data.} We obtain land use data from the Verisk National Land Use Database \footnote{\url{https://digimap.edina.ac.uk/roam/map/verisk}}, which provides high-resolution classification of land use types. The dataset includes ten primary land use categories. To create the evaluation dataset, we sample locations with a 200-meter radius buffer, ensuring balanced representation across categories.

\textbf{Socioeconomic data.} We obtain socioeconomic data from the Office for National Statistics (ONS) 2021 Census\footnote{\url{https://www.ons.gov.uk/}}, specifically the National Statistics Socioeconomic Classification (NS-SeC). This dataset provides a detailed classification of socioeconomic status based on employment type, occupational hierarchy, and educational attainment. The data are aggregated at the Lower-layer Super Output Area (LSOA) level, encompassing 4,994 LSOAs across London. Each LSOA contains proportions of 1000 to 3000 residents within different occupational classes.

\subsection{Baselines}

To assess the effectiveness of our enhanced model, CaLLiPer+,  we compare it against the following baselines:

\begin{itemize}
    \item \textbf{TF-IDF} \cite{sparck1972tfidf}: A term frequency-inverse document frequency model that represents each region based on the POI categories within it.
    \item \textbf{LDA} \cite{blei2003latent}: A probabilistic topic modeling approach that infers latent topics from POI distributions, capturing urban functional structures through topic-word distributions.
    \item \textbf{Place2Vec} \cite{yan2017itdl}: A spatial embedding model that learns representations of POIs based on their spatial co-occurrence, modeling functional similarity through a skip-gram framework.
    \item \textbf{Doc2Vec} \cite{niu2021delineating}: A document embedding approach that treats urban regions as documents composed of POI categories, learning region representations through unsupervised learning.
    \item \textbf{SPPE} \cite{huang2022sppe}: A semantics-preserving POI embedding method that captures spatial co-occurrence patterns and topological structures of POIs through a graph-based approach.
    \item \textbf{Space2Vec} \cite{mai2020iclr}: A geospatial representation learning model that encodes locations through positional encoding and neural networks, learning embeddings directly from spatial coordinates.
    \item \textbf{CaLLiPer} \cite{wang2025multi}: The original multimodal contrastive learning model, which encodes POI categories as textual descriptions but does not incorporate POI names.
\end{itemize}

\subsection{Downstream Tasks and Evaluation Metrics}
\label{subsec:downstream}
We evaluate the learned spatial representations on LUC and SDM tasks. To systematically analyze the effectiveness of the learned embeddings, we employ two types of downstream models: (1) a linear model, implemented as a single-layer neural network, testing the raw expressiveness of the embeddings, and (2) a nonlinear model, implemented as a multi-layer perceptron (MLP) with a single hidden layer to capture more complex relationships.

\textbf{Land use classification} is a multi-class classification task that predicts the land use type of a given spatial unit based on its learned representation. We train classifiers using both a linear model and a nonlinear model and evaluate performance using:
\begin{itemize}
    \item Precision, recall, and F1 score: These metrics are macro-averaged across classes, providing a balanced evaluation of classification performance. Higher values indicate better performance.
\end{itemize}

\textbf{Socioeconomic status distribution mapping} is a regression task that estimates the occupational composition of urban regions using the learned embeddings. The model predicts the proportion of residents in different socioeconomic categories at the LSOA level. We train both a linear model and a nonlinear model to compare their effectiveness. Performance is evaluated using:
\begin{itemize}
    \item L1 distance: Measures the absolute difference between predicted and actual socioeconomic distributions.
    \item Chebyshev distance: Captures the maximum absolute deviation between predicted and actual distributions.
    \item Kullback-Leibler (KL) divergence: Evaluates the difference between the predicted and actual probability distributions, indicating how well the model captures the socioeconomic structure.
\end{itemize}

By testing the embeddings across both classification and regression tasks, and using both linear and nonlinear models, we assess their generalizability and effectiveness in capturing the information of urban environments.

\subsection{Implementation Details}

All models were implemented using PyTorch and trained on a machine equipped with an NVIDIA A6000 GPU. The text encoder was based on Sentence-BERT by default, which processed the enriched POI descriptions. The spatial encoder followed the same architecture as in CaLLiPer \cite{wang2025multi}, using a fully connected residual network with 128-dimensional embeddings. The training process adopted a grid search approach to tune hyperparameters, resulting in a batch size of 128, a learning rate of 0.0001, and a temperature parameter of 0.07. The optimizer was Adam. The models were trained for 100 epochs with early stopping based on validation loss, and each downstream task experiment was repeated five times with different random seeds to ensure robustness. The reported results represent the mean performance across these runs.

\subsection{Location Retrieval}
\label{sec:location retrieval}
We observe the model’s ability to retrieve urban concepts based on semantic queries. This task shows how well the learned embeddings capture urban concepts by matching textual embeddings to spatial embeddings. 

Given a natural language query, we compute its embedding using a pretrained language model. We use two text encoding approaches: (1) a Sentence-Transformers model (all-MiniLM-L6-v2), which generates sentence embeddings via mean pooling over contextualized token embeddings, and (2) an OpenAI GPT-based embedding model (text-embedding-3-small), which produces a high-dimensional representation of the query and is subsequently projected into a 128-dimensional space for compatibility with the learned spatial embeddings. 

The model then retrieves the most relevant locations by computing cosine similarity between the query embedding and the location embeddings of urban regions. To assess retrieval effectiveness, we visualize the top-ranked locations using geospatial maps, highlighting areas with the highest similarity to the input query. 

\subsection{Ablation Study}

To evaluate the impact of different semantic components and text encoders, we conduct an ablation study with four model variants:

\begin{itemize}
    \item \textbf{CaLLiPer+ GPT}: A variant that replaces the sentence transformer with GPT (text-embedding-3-small), examining the effect of a text embedding from LLM. For fairness, we only use the first 384 dimensions of the text embedding, which is the same as the default sentence transformer.
    \item \textbf{CaLLiPer+}: The default enhanced model that integrates both POI names and types, using a sentence transformer (all-MiniLM-L6-v2).
    \item \textbf{CaLLiPer+ w/o type}: A variant that removes POI types, using only POI names for textual representation.
    \item \textbf{CaLLiPer}: A variant that excludes POI names and relies only on POI types, which is the original CaLLiPer.
\end{itemize}

We evaluate these models on the LUC and SDM tasks. The primary metrics used are F1 score for classification and KL divergence for regression-based analysis. The results are summarized in Figure~\ref{fig:ablation_study}.

\section{Results and Analysis}
\label{sec:results}

\subsection{Performance on Downstream Tasks}
\label{subsec:results_downstream}

\begin{table}[t]
\caption{Performance comparison on the LUC task. The best and second-best performances are marked in \textbf{bold} and \underline{underlined}, respectively.  For better readability, all metrics are scaled by a factor of $10^{2}$.}
\label{tab:luc_res}
\centering
\resizebox{\textwidth}{!}{ 
\begin{tabular}{l|ccc|ccc}
\toprule
\multirow{2}{*}{Model} & 
\multicolumn{3}{c|}{Linear} & \multicolumn{3}{c}{MLP} \\
 & Precision ↑ & Recall ↑ & F1 Score ↑ & Precision ↑ & Recall ↑ & F1 Score ↑ \\
\midrule
Random & 9.6 ± 0.7 & 10.3 ± 0.5 & 9.7 ± 0.5 & 8.8 ± 1.3 & 10.3 ± 0.3 & 9.0 ± 0.3 \\
TF-IDF & 31.5 ± 0.4 & 32.2 ± 0.2 & 31.3 ± 0.3 & 31.8 ± 0.6 & 33.3 ± 0.5 & 31.7 ± 0.6 \\
LDA & 30.8 ± 0.3 & 29.1 ± 0.2 & 28.4 ± 0.2 & 31.5 ± 1.1 & 30.4 ± 0.7 & 29.2 ± 0.9 \\
Place2Vec & 30.9 ± 0.8 & 26.1 ± 0.7 & 26.3 ± 0.7 & 35.1 ± 1.2 & 32.7 ± 1.0 & 32.4 ± 1.2 \\
Doc2Vec & 32.4 ± 0.4 & 28.2 ± 0.1 & 28.0 ± 0.1 & 34.9 ± 0.9 & 33.8 ± 0.5 & 32.7 ± 0.6 \\
SPPE & 30.5 ± 0.4 & 27.0 ± 0.2 & 26.6 ± 0.2 & 34.5 ± 0.9 & 32.9 ± 0.7 & 32.2 ± 0.5 \\
HGI & 33.0 ± 0.5 & 30.0 ± 0.6 & 29.9 ± 0.6 & 33.6 ± 0.5 & 32.0 ± 0.9 & 31.6 ± 0.7 \\
Space2Vec & 28.6 ± 0.6 & 28.5 ± 0.8 & 27.4 ± 0.7 & 29.6 ± 0.6 & 28.9 ± 0.5 & 27.8 ± 0.3 \\
CaLLiPer & 36.5 ± 0.6 & 35.3 ± 0.2 & 34.6 ± 0.3 & 37.7 ± 0.8 & 35.5 ± 0.8 & 34.6 ± 0.8 \\
\midrule
CaLLiPer+ & \underline{37.5 ± 0.7} & \underline{35.5 ± 0.5} & \underline{35.2 ± 0.6} & \underline{40.0 ± 0.4} & \underline{36.0 ± 0.5} & \underline{36.6 ± 0.5} \\ CaLLiPer+GPT \normalsize & \textbf{40.5 ± 0.6} & \textbf{36.7 ± 0.2} & \textbf{36.8 ± 0.3} & \textbf{41.3 ± 0.7} & \textbf{37.8 ± 0.4} & \textbf{37.6 ± 0.3} \\
\bottomrule
\end{tabular}
}
\end{table}

\begin{table}[t]
\caption{Performance comparison on the SDM task. The best and second-best performances are marked in \textbf{bold} and \underline{underlined}, respectively. For better readability, all metrics are scaled by a factor of $10^{2}$.}
\resizebox{\textwidth}{!}{ 
\begin{tabular}{l|ccc|ccc}
\toprule
\multirow{2}{*}{Model} & \multicolumn{3}{c|}{Linear} & \multicolumn{3}{c}{MLP} \\
 & L1 ↓ & Chebyshev ↓ & KL ↓ & L1 ↓ & Chebyshev ↓ & KL ↓ \\
\midrule
Random & 30.31 ± 0.03 & 9.25 ± 0.01 & 7.73 ± 0.01 & 31.40 ± 0.22 & 9.55 ± 0.11 & 8.21 ± 0.14 \\
TF-IDF & 24.79 ± 0.04 & 7.43 ± 0.01 & 5.36 ± 0.01 & 24.36 ± 0.15 & 7.28 ± 0.05 & 5.20 ± 0.04 \\
LDA & 26.14 ± 0.01 & 7.84 ± 0.00 & 5.87 ± 0.00 & 25.85 ± 0.14 & 7.77 ± 1.12 & 5.80 ± 0.72 \\
Place2Vec & 23.47 ± 0.09 & 6.94 ± 0.02 & 4.81 ± 0.02 & 22.81 ± 0.06 & 6.81 ± 0.01 & 4.61 ± 0.02 \\
Doc2Vec & 24.01 ± 0.07 & 7.15 ± 0.02 & 4.99 ± 0.02 & 23.10 ± 0.19 & 6.89 ± 0.06 & 4.75 ± 0.08 \\
SPPE & 24.32 ± 0.16 & 7.24 ± 0.06 & 5.11 ± 0.06 & 23.63 ± 0.19 & 7.04 ± 0.06 & 4.91 ± 0.07 \\
HGI & 23.28 ± 0.08 & 6.93 ± 0.02 & 4.79 ± 0.03 & 22.73 ± 0.05 & 6.80 ± 0.02 & 4.60 ± 0.02 \\
Space2Vec & 25.13 ± 0.15 & 7.56 ± 0.04 & 5.65 ± 0.06 & 23.55 ± 0.20 & 7.12 ± 0.09 & 5.00 ± 0.08 \\
CaLLiPer & 21.63 ± 0.04 & 6.55 ± 0.05 & 4.26 ± 0.01 & 20.52 ± 0.14 & 6.24 ± 0.03 & 3.90 ± 0.06 \\
\midrule
CaLLiPer+ & \underline{20.87 ± 0.02} & \underline{6.35 ± 0.01} & \underline{3.98 ± 0.01} & \underline{19.85 ± 0.19} & \underline{6.02 ± 0.06} & \underline{3.63 ± 0.07} \\
CaLLiPer+GPT & \textbf{20.26 ± 0.03} & \textbf{6.09 ± 0.01} & \textbf{3.74 ± 0.01} & \textbf{19.38 ± 0.02} & \textbf{5.83 ± 0.04} & \textbf{3.47 ± 0.01} \\
\bottomrule
\end{tabular}
}
\label{tab:sdm_res}
\end{table}
Tables~\ref{tab:luc_res} and~\ref{tab:sdm_res} summarize the results for LUC and SDM tasks. Across both tasks, multimodal contrastive learning models outperform traditional methods, demonstrating the effectiveness of integrating spatial and semantic information. Baseline models such as TF-IDF and LDA rely on aggregated POI type distributions within regions, limiting their ability to capture fine-grained relationships between locations. While methods like Place2Vec and Doc2Vec improve upon this by incorporating spatial co-occurrence structures, their reliance on unsupervised embedding techniques without explicit spatial-semantic alignment leads to weaker performance. In contrast, CaLLiPer and its extensions, which align POI-based textual representations with spatial coordinates, consistently achieve better results, confirming the advantages of multimodal contrastive learning.

Additionally, CaLLiPer+ achieves superior and more stable performance across all metrics. In LUC, CaLLiPer+ consistently outperforms the original CaLLiPer model, achieving higher precision, recall, and F1 scores across both linear and MLP classifiers. This demonstrates that integrating POI names alongside type-based descriptions enriches the model’s semantic understanding of urban space, allowing for better land use classification. A similar trend is observed in SDM, where CaLLiPer+ further reduces errors across all three evaluation metrics, suggesting that POI names provide valuable contextual information for modeling socioeconomic distributions. Notably, CaLLiPer+ GPT achieves the best performance across both tasks, reinforcing the importance of using more powerful text encoders for spatial representation learning.

Third, the improvements observed with MLP over the linear model suggest that the learned embeddings still contain complex, non-linear relationships that can be further leveraged by downstream tasks. While baseline models such as TF-IDF and LDA show limited gains with MLP, indicating that their representations are mostly exhausted by simple classifiers, CaLLiPer-based models still exhibit a more notable performance boost. CaLLiPer+ effectively aligns spatial and semantic information, and the embeddings still retain structured patterns that require more expressive models to fully exploit, highlighting the depth and richness of the learned representations.

These findings highlight the advantages of incorporating both POI names and stronger text embedding models for geospatial representation learning, improving the model’s ability to capture complex urban semantics across diverse tasks.
\subsection{Location Retrieval}
\label{subsec:retrieval}
\begin{figure}[t]
    \centering
    \includegraphics[width=1\linewidth]{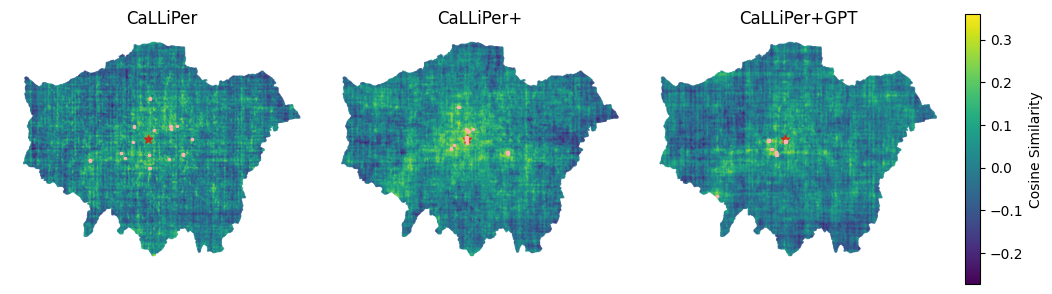}
    \caption{Similarity map for "The National Gallery." The red star is the actual location of the target, and the yellow points are the top 30 similar locations.}
    \label{fig:retrieval_1}
\end{figure}

\begin{figure}[t]
    \centering
    \includegraphics[width=0.7\linewidth]{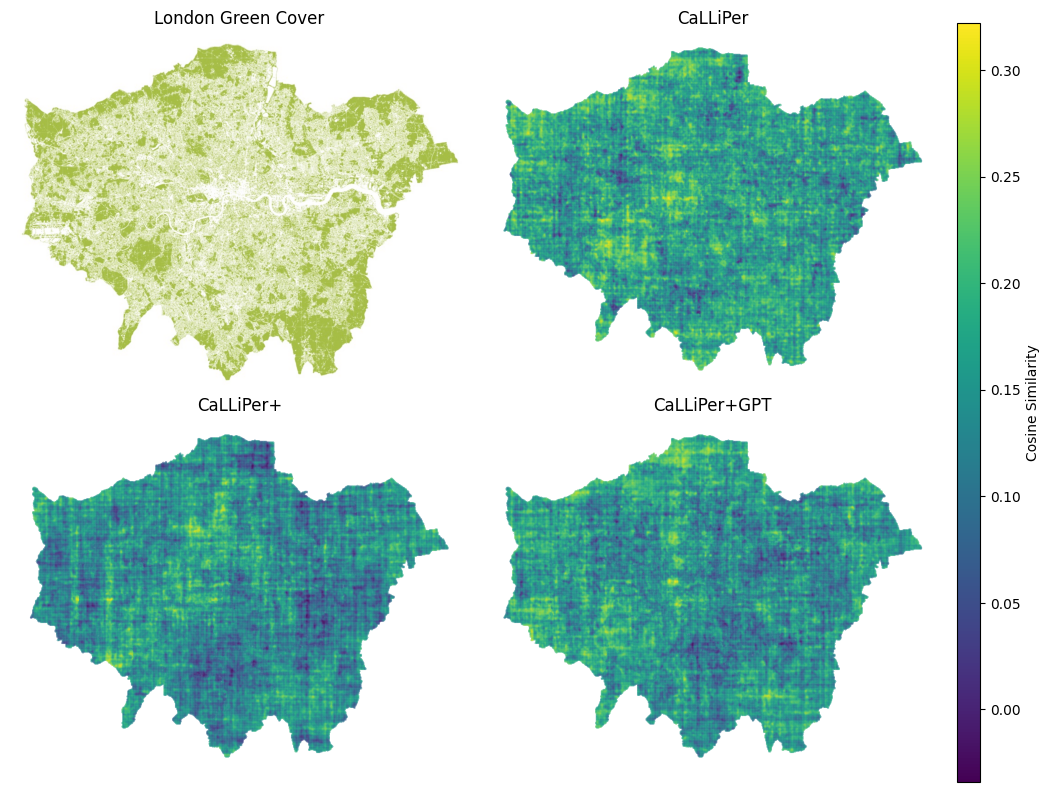}
    \caption[Similarity map for "A place of park or green cover."]{Similarity map for "A place of park or green cover." The ground truth is based on green cover data from London DataStore \footnotemark.}
    \label{fig:retrieval_2}
    
\end{figure}

Location retrieval evaluates the model’s ability to associate spatial embeddings with meaningful semantic queries, including specific place names and abstract urban concepts. The results, shown in Figures~\ref{fig:retrieval_1} and~\ref{fig:retrieval_2}, illustrate how different model variants respond to retrieval tasks.

First, using POI names directly for retrieval demonstrates that including POI names in the text encoder significantly improves the model’s ability to locate specific places. In Figure~\ref{fig:retrieval_1}, models that incorporate POI names (CaLLiPer+ and CaLLiPer+GPT) produce more precise and concentrated retrieval results compared to the original CaLLiPer model, which relies solely on categorical types. The use of a more powerful text encoder, such as GPT embeddings in CaLLiPer+GPT, further enhances localization, leading to more accurate spatial responses.

Second, for high-level conceptual retrieval, such as identifying regions characterized by abstract urban concepts (e.g., green cover), the inclusion of POI names introduces both benefits and challenges. As seen in Figure~\ref{fig:retrieval_2}, models that incorporate POI names sometimes exhibit increased dispersion in similarity scores when handling broad, high-level concepts. This suggests that when the model’s semantic understanding is insufficient, in such cases, additional name-based details can introduce ambiguity. However, when equipped with a more advanced text encoder (e.g., CaLLiPer+GPT), the model can effectively utilize this additional semantic information to establish clearer distinctions between different urban functions, demonstrating improved conceptual retrieval. This improvement can be attributed to GPT’s ability to capture hierarchical urban concepts and their interconnections, enabling a more nuanced understanding of spatial semantics.

Overall, our results highlight the benefits of integrating POI names in location retrieval. Name-enhanced models improve direct place retrieval and, with sufficiently strong text encoders, also facilitate better discrimination of abstract spatial concepts.

\footnotetext{\url{https://apps.london.gov.uk/green-cover}}

\subsection{Ablation Study Results}
\begin{figure}[t]
    \centering
    \includegraphics[width=0.75\linewidth]{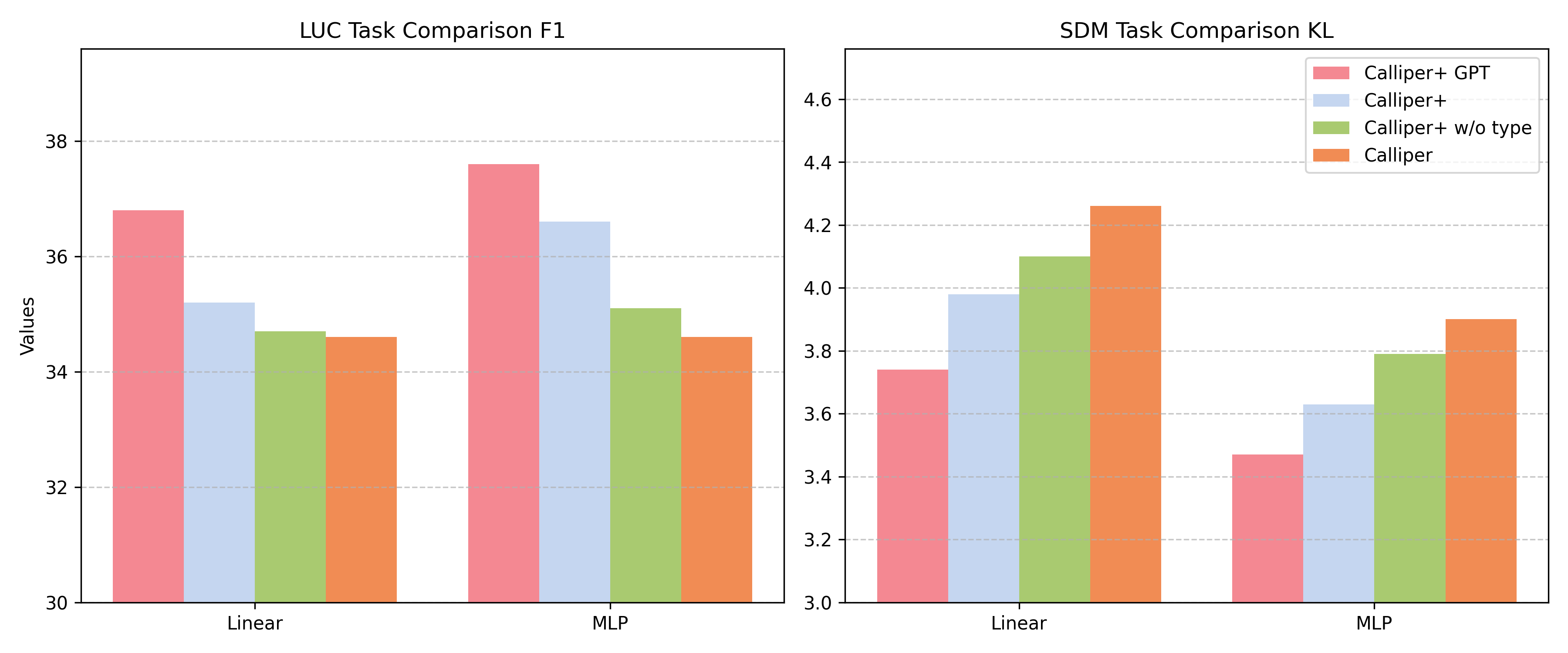}
    \caption{Ablation study results comparing model variations across LUC and SDM tasks. The left plot shows F1 score ↑ performance on LUC, while the right plot presents KL divergence ↓ results for SDM. All metrics are scaled by a factor of $10^{2}$.}
    \label{fig:ablation_study}
\end{figure}

Figure~\ref{fig:ablation_study} presents the results of our ablation study. Both POI names and types contribute to improving downstream tasks, as seen from the superior performance of CaLLiPer+ compared to CaLLiPer and CaLLiPer+ w/o type. This suggests that combining both sources of semantic information leads to more informative spatial representations. 

Interestingly, even when POI types are removed (CaLLiPer+ w/o type), the model still outperforms CaLLiPer, indicating that POI names carry richer and more discriminative semantic details than type labels alone. This highlights the potential of leveraging fine-grained textual information like POI names in spatial embedding models.

Moreover, using a stronger text encoder (CaLLiPer+ GPT) further improves results across both tasks. The enhanced semantic representation from a large language model allows for a better understanding of the text concepts in urban semantics, reinforcing the importance of high-quality embeddings in geospatial contrastive learning.

\section{Discussion and Conclusion}
\label{sec:discussion}
We explore the impact of integrating POI names into multimodal contrastive learning for spatial representation. By extending the CaLLiPer framework to incorporate both POI types and names, we introduce CaLLiPer+, which enhances the semantic richness of location embeddings. Our experiments across land use classification, socioeconomic status distribution mapping, and location retrieval reveal key insights into the role of enriched textual descriptions in geospatial learning.

\textbf{Effectiveness of POI names in spatial representation.}
The combining of POI names with types in multi-modal contrastive learning improves downstream task performance consistently. POI names provide more specific and context-aware semantic signals, capturing fine-grained distinctions that categorical types alone may overlook. This effect is particularly evident in retrieval tasks, where name-enhanced models demonstrate greater precision in identifying specific locations.

\textbf{Impact of text encoder strength.}
Using more advanced text embeddings, such as those from GPT-based models, further refines spatial representation. The CaLLiPer+ GPT model consistently outperforms others, suggesting that stronger language models contribute to a deeper understanding of urban semantics. This aligns with findings in location retrieval, where better text embeddings enable clearer conceptual differentiation, especially for high-level concepts.

\textbf{Limitations and future work.}  
The quality of spatial embeddings relies on the density and distribution of POIs across different urban areas. Regions with too sparse POI coverage may lead to less informative representations, limiting generalizability. Also, the information beyond the semantics still needs to be explored. Future work should incorporate additional modalities such as road networks, street-view imagery, and mobility patterns to enrich spatial information. Additionally, while our current downstream tasks provide initial validation, further research should explore a wider range of urban analytics applications and develop task-specific models that better leverage the structure of learned embeddings for improved adaptability and performance.

\textbf{Conclusion.}
This work demonstrates that incorporating POI names into geospatial contrastive representation learning leads to improved performance in multiple urban analytics tasks. By aligning spatial and semantic information more effectively, CaLLiPer+ provides a more detailed and context-aware model for understanding urban environments. The effectiveness of semantic information highlights the potential of using pretrained multimodal models to generate enriched spatial embeddings in advancing urban intelligence.


\bibliography{references}

\appendix

\end{document}